\documentstyle[11pt]{article}
\begin{document}

\begin{titlepage}
\begin{flushright}
hep-th/9704017\\
\end{flushright}
\vspace{.5cm}

\begin{center}
{\LARGE A Possible IIB Superstring Matrix Model with Euler Characteristic
\\[.4cm]
and a Double Scaling Limit}\\
\vspace{1.2cm}
{\large C. F. Kristjansen\footnote{E-mail: kristjan@alf.nbi.dk} and 
P. Olesen\footnote{E-mail: polesen@nbi.dk}} 
\vspace{24pt}

{\it The Niels Bohr Institute,}
\\ {\it  Blegdamsvej 17, DK 2100 Copenhagen \O, Denmark}
\end{center}
\vskip 0.9 cm
\begin{abstract}
We show that a recently proposed Yang-Mills matrix model with an auxiliary
field, which is a candidate for a non-perturbative description of type IIB 
superstrings, captures the Euler characteristic of moduli space of Riemann 
surfaces. This happens at the saddle point for the Yang-Mills field. It turns 
out that the large-$n$ limit in this matrix model corresponds to a double 
scaling limit in the Penner model. 
\end{abstract}
\end{titlepage}
\setcounter{page}{1}
\renewcommand{\thefootnote}{\arabic{footnote}}
\setcounter{footnote}{0}
The conjecture by Banks, Fischler, Shenker, and Susskind \cite{BFSS96} that M 
theory is an $n=\infty$ matrix model, has induced a large activity towards
a non-perturbative formulation of type IIB superstrings by matrix models.
It has been proposed by Ishibashi, Kawai, Kitazawa, and Tsuchiya (IKKT)
\cite{japan} that the large-$n$ reduced model of ten-dimensional super 
Yang-Mills theory can be considered as such a formulation, provided the matrix
size, $n$, is considered as a dynamical variable to be summed from $n=0$ to 
$n=\infty$ in the partition function. It was recently proposed \cite{nbi} that
this sum over all $n$, which is somewhat foreign to the $n=\infty$ philosophy,
should be replaced by an auxiliary matrix field $Y^i_j$, and the 
large-$n$ limit taken, thereby avoiding introducing small-$n$ matrices. The
partition function is then given by \cite{nbi} 
\begin{equation}
Z_{\rm NBI}=\int {\cal D}A_\mu{\cal D}Y~e^{-S_\epsilon}=C\int {\cal D}[A_\mu]
e^{-S_{\rm NBI}},
\label{1}
\end{equation}
where $C$ is a constant, and the ``dielectric'' action $S_\epsilon$ is given by
\begin{equation}
S_\epsilon=-\frac{\tilde{\alpha}}{4}~{\rm Tr}\left(Y^{-1}[A_\mu,A_\nu]^2\right)
+\tilde{\beta}~{\rm Tr}~Y+(n-\frac{1}{2})~{\rm Tr}~\ln Y,
\label{2}
\end{equation}
and where the non-Abelian (strong field) Born-Infeld action $S_{\rm NBI}$ is 
given by
\begin{equation}
S_{\rm NBI}=-\sqrt{\tilde{\alpha}\tilde{\beta}}~{\rm Tr}
\sqrt{-[A_\mu,A_\nu]^2}.
\label{3}
\end{equation}
Also, the measure $[{\cal D}A_\mu]$ is given by
\begin{equation}
[{\cal D}A_\mu]={\cal D}A_\mu~\prod_{i<j}(z_i+z_j)^{-1},
\label{4}
\end{equation}
where the $z$'s are the eigenvalues of the commutator $-[A_\mu,A_\nu]^2$.

In writing these results we have ignored the fermionic parts of these actions
(as well as the fermionic functional integrations), because they play no
role in the following. The NBI action (\ref{3}) allows one to pass to
the Nambu-Goto form of superstrings, provided the $n\rightarrow\infty$
limit can be performed inside various sums ($\approx$ absence of tachyon
singularities). We refer to \cite{nbi} for the complete actions and a
discussion of the limits involved in obtaining the Nambu-Goto string square 
root from $S_{\rm NBI}$.

It is worth emphasizing that the unconventional term Tr ln $Y$ in eq. (\ref{2})
is needed in order to get the square root action (\ref{3}). Likewise
the coefficient $n-1/2$ is crucial for the $Y$-integral
to give eq. (\ref{3}).

In this paper we address the question concerning the meaning of the
auxiliary field $Y$ in (\ref{2}). We shall show that {\sl the physical
interpretation of $Y$ is that it captures the Euler characteristic of moduli
space of Riemann surfaces}. It turns out, rather surprisingly, that the limit
$n\rightarrow\infty$ in the $A_\mu$ saddle point action, arising from
(\ref{2}), automatically implies a double scaling 
limit~\cite{dsl}, so that the 
partition function (\ref{1}) contains the sum over {\rm all} (virtual) Euler 
characteristic of moduli space of genus $g$. These results are closely related
to works on the Penner model \cite{penner,dv,CDL91,Tan91,akm}. The 
appearance of the ``automatic double scaling'' limit is rather encouraging
from the point of view that the partition function (\ref{1}) should have
something to do with superstrings.

We shall now look for a saddle point of the action (\ref{2}) in $A_\mu$. As
shown in ref. \cite{nbi}, this implies
\begin{equation}
[A_\mu,\{Y^{-1},[A_\mu,A_\nu]\}]=0.
\label{5}
\end{equation}
Here $\{a,b\}$ denotes the anti-commutator of $a$ and $b$. The equation of
motion (\ref{5}) differs from the one considered by IKKT \cite{japan}
by the presence of the $Y$-field. Similarly to IKKT we can find a solution by
using that the $A_\mu$-field commutes with the unit matrix, and hence
\begin{equation}
\{Y^{-1},[A_\mu,A_\nu]\}^i_j=2im_{\mu\nu}~\delta^i_j,~{\rm with}~m_{\mu\nu}
=-m_{\nu\mu},
\label{6}
\end{equation}
where $m_{\mu\nu}$ is a matrix with respect to the space indices, examples of
which are given by IKKT \cite{japan} for D-strings. The solution of
(\ref{6}) is given by
\begin{equation}
[A_\mu,A_\nu]^i_j=im_{\mu\nu}~Y^i_j.
\label{7}
\end{equation}
{}From eq. (\ref{7}) it follows that the action (\ref{2}) at the saddle point
is given by
\begin{equation}
S^{\rm saddle}_\epsilon=na~{\rm Tr}~Y+(n-\frac{1}{2})~{\rm Tr}~\ln~Y,
\label{8}
\end{equation}
where the parameter $a$ is defined by
\begin{equation}
a=(\tilde{\beta}+\tilde{\alpha}~m_{\mu\nu}^2/4)/n.
\label{9}
\end{equation}
The parameters $\tilde{\alpha}$ and $\tilde{\beta}$ should be positive
but are not otherwise fixed by the work
in ref. \cite{nbi}. In order to have a non-trivial large-$n$ saddle point we 
need to assume that
\begin{equation}
\tilde{\alpha}=n\alpha~{\rm and}~\tilde{\beta}=n\beta,
\label{10}
\end{equation}
where the quantities without tildes are of order one. This is the reason for 
defining $a$ as in (\ref{9}). It should be noted that  $a$ is positive.

An important point is that taking $\tilde{\alpha}$ and $\tilde{\beta}$
of order $n$ is not equivalent to taking the usual classical limit in
string theory, i.e. $g_s=0$. The limit $g_s=0$  appears for 
$\alpha,\beta\rightarrow\infty$ \cite{japan}, and hence the term Tr ln $Y$ is 
ignored in this limit. Here we keep $\alpha$ and $\beta$ finite, and thus
 the string coupling does not vanish. 
Another way of seeing this is by noticing that at the saddle point
the $A_\mu$-fields satisfy eqn.\ (\ref{7}) where the auxiliary field $Y$
is arbitrary, subject only to the constraints coming from integrating
over $Y$. More precisely, at the saddle point the commutator
$\left[A_{\mu},A_{\nu}\right]$ has a non-trivial distribution of
eigenvalues, determined by the distribution of eigenvalues for the
matrix $Y$. The distribution of eigenvalues of the matrix $Y$, on the
other hand, is determined by the saddle point action~(\ref{8}). The
matrix model defined by the action~(\ref{8}) is a well-understood
model and, as we shall see, we are led to the conclusion that the
field $Y$ is a book-keeping field which allows us to distinguish
between contributions from different genera. 
 
 The value of the partition 
function in the saddle point is given by the expression
\begin{equation}
Z^{\rm saddle}_{\rm NBI}={\rm const.}\int {\cal D}Y~\exp[-n~{\rm Tr}
(a~Y+t~\ln~Y)]
\times {\cal G}
\label{11}
\end{equation}
where $\cal G$ represents the terms resulting from the gaussian integration 
over the fluctuations of the $A$-fields. In the following we shall ignore
the sub-dominant factor $\cal G$. It follows from eq. (\ref{2}) that the 
parameter $t$ is 
\begin{equation}
t=1-1/2n.
\label{12}
\end{equation}

Now the crucial point is that the action (\ref{11}) is of the Penner type
\cite{penner}. We refer to refs. \cite{dv,CDL91,Tan91,akm} for a detailed
discussion of this model. It should also be noticed that for the $t$-value
(\ref{12}) the $Y$-integral in (\ref{11}) diverges\footnote{In doing the 
exact $Y$-integral in eq.(\ref{1}) this problem
does not occur \cite{nbi}, because of the term containing $Y^{-1}$ in the
action $S_\epsilon$ in eq. (\ref{2}). This integral can be performed
restricting the $Y$-integration to run over positive eigenvalues of the 
Hermitian matrix $Y$.}. Therefore we consider
$t$ first as a free parameter to be used in an analytic continuation.
Ultimately $t$ will reach the value given in (\ref{12}).

The Penner model is given by the following matrix integral
\begin{equation}
Z_{\rm Penner}=e^{ F}=\exp(-n^2t)\int d\phi \exp\{-nt~{\rm Tr}
\left(\ln \phi -\phi\right)\}.
\label{35}
\end{equation}
To begin with this model is defined only for negative $t$ (similarly
to eq. (\ref{11})), but by analytical
continuation one can extend its definition also to $t>0$. There are two ways
of evaluating the integral in (\ref{35}): Either one can use the method of
orthogonal polynomials with $t$ less than zero \cite{dv} and obtain the
result for all $n$ and for positive $t$ by analytic continuation, 
or one can
evaluate it order by order in $1/n$ using a saddle point 
method~\cite{CDL91,Tan91,akm}. In the last 
method the integration is deformed into the complex plane, and in general the
matrix $Y$ will have an eigenvalue distribution with complex eigenvalues,
although the matrix $Y$ originally is 
Hermitian~\cite{CDL91,akm}. Of course, the non-real eigenvalues
occur in complex conjugate pairs, so the resulting free energy is real. 

Thus, in the language
of the spectral density the region $t>0$ is characterised by the eigenvalues
of the matrix $\phi$ not being restricted to the real axis but living on
some curve in the complex plane. 
Examples of such curves for various negative values of $t$ are
shown in figs. 3--8 in ref.
\cite{akm}. 
Using any of the methods in ref. \cite{dv} or \cite{CDL91,Tan91,akm},
it turns out that the (analytically continued) Penner model has a critical 
point at $t=t_c=1$. In the saddle point method,  
at this point the 
endpoints of the support of the eigenvalue 
distribution coalesce and the support of the eigenvalue distribution forms
a closed loop in the complex plane~\cite{CDL91,akm}. 
In the vicinity of the critical point
$t=t_c=1$ one can define a double scaling limit (in both approaches \cite{dv}
and \cite{CDL91,Tan91,akm})
\begin{equation}
t\rightarrow (t_c)_-,~ n\rightarrow \infty,~{\rm and}~ 
\mu=(t_c-t)\, n={\rm  fixed},
\label{scaling}
\end{equation}
so that the contribution to
the free energy from surfaces of genus $g$ reads\footnote{For a discussion
of the genus zero free energy $F_0$, and $F_1$ see refs. 
\cite{dv} and \cite{CDL91,Tan91,akm}.}
\cite{dv}
\begin{equation}
F=F_0+F_1+\sum_{g=2}^{\infty}n^{2-2g}\,f_g,
\label{Fg}
\end{equation}
with
\begin{equation}
f_g=\frac{B_{2g}}{2g(2g-2)}(t_c-t)^{2-2g},~~g>1,
\label{Fgg}
\end{equation}
where $\{B_{2g}\}$ are the Bernoulli numbers and 
$\chi_g=\frac{B_{2g}}{2g(2g-2)}$ is the virtual Euler characteristic of moduli
space of Riemann surfaces of genus $g$.
In the double scaling limit
(\ref{scaling}) the expansion (\ref{Fg}) thus takes the form
\begin{equation}
F(\mu)=F_0(\mu)+F_1(\mu)+\sum_{g=2}^{\infty}\frac{B_{2g}}{2g(2g-2)}\mu^{2-2g}.
\label{38}
\end{equation}
As usual, the merit of the double scaling limit is that all genera
 are included.

In the double scaling limit defined in eq. (\ref{scaling}), using the
approach in ref. \cite{CDL91,akm}, the endpoints
of the support of the eigenvalue distribution $x$ and $y$ 
lie on the real axis and are separated by an
infinitesimal distance of the order 
$(1-t)^{1/2}\sim O(1/\sqrt{n})$. The support of the eigenvalue distribution is 
a curve, starting out at $x$ and running along the real axis towards $y$ but
making a loop in the complex plane encircling the origin before
actually reaching the point $y$~\cite{akm}. 
 In the approach of
Distler and Vafa \cite{dv}, the integral is anyhow evaluated exactly in
the region where it converges (integration over only the positive eigenvalues
of the hermitian matrix $\phi$), and then analytically continued.

Let us now return to the matrix model given by eq. (\ref{1}), and consider the
saddle point value (\ref{11}), which is clearly of the Penner type. Here it
should be remarked that if we change the coupling constant in front of the 
linear term in the Penner action in (\ref{35}), (keeping its sign fixed), 
this only changes the partition function by an irrelevant constant
and does not alter the position of the critical point or the
behaviour~(\ref{Fg}) and~(\ref{Fgg}). 
Therefore the results discussed above can be taken over
without any changes. Due to eq. (\ref{12}) we see that for this 
model~\footnote{Since the critical 
point $t_c=1$ appears, this indicates a connection between the
``dielectric'' matrix model \cite{nbi} and $c=1$ matter coupled to
two-dimensional gravity.}  
\begin{equation}
t_c=1,~~t_c-t=1/2n,~~{\rm i.e.}~~\mu=(t_c-t)n=1/2={\rm fixed}.
\label{36}
\end{equation}
Therefore the matrix model (\ref{1}) {\sl automatically} satisfies the
double scaling relations (\ref{scaling}) in the limit $n\rightarrow\infty$
with the ``cosmological constant'' $\mu$ equal to
1/2. Thus, information about higher genera contributions is encoded
in the ``dielectric'' matrix model of ref. \cite{nbi} in the limit 
$n\rightarrow \infty$. In this sense it is
reasonable to assume that the auxiliary field $Y$ is really a substitute for
the summation over $n$ in the IKKT approach \cite{japan}.

The appearance of the Bernoulli numbers $B_{2g}$ with positive coefficients
shows that the genus expansion of the partition function is not Borel
summable. It is well known that the same is true for the genus
expansion in string theory \cite{gross}. 

It should be mentioned that the matrix model (\ref{1}) also knows about
the Euler characteristic $\chi_{g,q}$ 
of the moduli space of Riemann surfaces of 
genus $g$ with $q$ punctures,
\begin{equation}
\frac{{\partial}^qF}{{\partial}\mu^q}=q!\sum \chi_{g,q}\,\mu^{2-2g-q}.
\end{equation}
This was shown for the Penner model in ref. \cite{dv}, and
can be repeated without any change for saddle point value of the model 
(\ref{1}) exhibited in eq. (\ref{11}). 

We end by a few remarks: The results obtained above
show that the matrix model proposed in ref. \cite{nbi} encodes some
highly non-trivial information about Riemann surfaces, which is 
crucial for strings. In particular, information about higher genera surfaces is
encoded already in the saddle point action, a feature which must be ascribed
to the non-perturbative character of the model. It is interesting to note
that the partition function (and the distribution of eigenvalues of the
auxiliary field) at the saddle point can be evaluated exactly. 
This opens the possibility of studying the model also beyond the saddle point
approximation.

\vspace{15pt}
\noindent
{\bf Acknowledgements}\hspace{0.3cm}
We thank Yu.\ Makeenko for interesting comments.
One of us (P.O.) in addition thanks A. Fayyazuddin for interesting discussions.

\end{document}